\RequirePackage{fix-cm}
\documentclass[smallextended]{svjour3}       
\smartqed  
\usepackage{graphicx}
\usepackage{geometry}                
\usepackage{amssymb,amsmath,mathtools,amsfonts}
\usepackage{epstopdf}
\usepackage{cite} 
\usepackage{color}

                                \newcommand{\Z}{\mathbb{Z}}
                                \newcommand{\eps}{\varepsilon} 
                                \newcommand{\mca}{\operatorname{MCA}} 
                                
\begin{document}

\title{The theory of games and microbe ecology
}


\author{Susanne Menden-Deuer         \and
        Julie Rowlett
}


\institute{S. Menden-Deuer \at
              Graduate School of Oceanography, University of Rhode Island, Narragansett, RI, USA \\
              \email{smenden@uri.edu} 
           \and
           J. Rowlett \at
              Department of Mathematics, Chalmers University and the University of Gothenburg, Sweden \\ \email{julie.rowlett@chalmers.se} 
}

\date{Received: date / Accepted: date}

\maketitle

\begin{abstract}
Using game theory we provide mathematical proof that if a species of asexually reproducing microbes does not possess maximum variability in competitive abilities amongst its individual organisms, then that species is vulnerable to replacement by competitors.  Furthermore, we prove that such maximally variable species are neutral towards each other in competition for limited resources; they coexist. Our proof is constructive:  given one species which does not possess maximum variability, we construct a species with the same (or lower) mean competitive ability which can invade, in the sense that its expected value in competition is positive whereas the expected value of the non-maximally variable species is negative. Our results point towards the mechanistic underpinnings for the frequent observations that (1) microbes are characterized by large intra-specific variability and that (2) the number of extant microbe species is very large.  
\end{abstract}

\section{Introduction}\label{intro}

The problem addressed here is the inexplicably high taxonomic and phenotypic diversity observed among and within microbes (e.g. plankton).   Microbes are the key engines of fundamental biological processes and show immense phenotypic variability within species in terms of physiological, demographic or morphological characteristics (i.e. traits). Henceforth, we call this characteristic \em intra-specific variability. \em   A fundamental conundrum arises because selection should favor individuals that maximize fitness (i.e. winner takes it all) and thus minimize intra-specific variation. This conundrum is well documented in the literature. Enabled by recent advances in measurement capacity,  key traits subject to selection show surprising diversity.  A non-exhaustive list of examples of highly variable traits includes:   elemental composition, morphology, physiology and behavior \cite{Boydetal2013, Moaletal1987,  RynearsonArmbrust2004, Fredricksonetal2011, Kiorboe2013, smdMontalbano2015}. Such intra-specific variability has even been documented in marine metazoa \cite{Morozovetal2013}. Moreover, demonstrable variation among clonal individuals has shown intra-specific variability enhances species dispersal rates, cell encounters and population distributions \cite{smd2010}. Recent empirical data confirm that variations in motility at the strain level manifest themselves in different population distributions \cite{Harveyetal2015}. Such intra-specific variability may enhance species survival and co-existence and has been suggested to be adaptive based on model simulations \cite{jrsi}. Observations of intra-specific variability in marine microbes are not restricted to artificial laboratory settings. Satellite observations show the distribution of a species of photosynthetic plankton persisting for months, over hundreds of kilometers along the North American west coast \cite{Duetal2011, Whiteetal2014}.  Laboratory experimentation on this plankton species documents intra-specific variability in physiology and movement behaviors \cite{smdMontalbano2015} that provide explanatory power for the species persistence documented with satellite observations.  Thus, in the laboratory and field tremendous intra-specific variability is demonstrable but difficult to explain theoretically.

Beyond the tremendous intra-specific physiological and behavioral variability discussed above, there is also an atonishing observable taxonomic diversity within marine microbes.  This vast diversity of species occupies all branches on the tree of life \cite{RynearsonPalenik2011}. Recent investigations have shown that although thousands of species have already been described and discovered, many if not most marine microbes are yet to be identified \cite{deVargasetal2015, Wordenetal2015}. Although the number of planktonic species is unknown, the likely number of species exceeds most if not all other groupings of organisms. Using data collected globally, researchers estimated that there may be around 150,000 eukaryotic plankton species \cite{deVargasetal2015}.  Moreover, if prokaryotic species are to be included, there are likely millions of species of marine microbes and their total number may not be knowable \cite{Ward2002}.

Here, we introduce a game theoretic model for microbes which aims to understand the mechanistic underpinnings that yield the observable diversity in species and behaviors. The model interpolates between  microscopic, individual level competition and macroscopic, species level survival.  Justified by observations, species are characterized by intra-specific variability, manifested in different distributions of their competitive abilities. To avoid trivial outcomes, the mean competitive ability is restricted to have a maximum value of $\frac 1 2$.  This can also be interpreted as assuming species are equally limited by energy and resources, when the species is assessed cumulatively. Species do vary in their trait distributions, which imparts different degrees of intra-specific variability from degenerate (= invariant) distributions to the uniform distribution with maximal variability. Species are represented by specific distributions that do not change over time but individuals' competitive abilities are chosen at random within each fixed cumulative distribution. Randomization of individual competitive abilities over time within a species represents competition in a dynamic system, where species are characterized by multiple physiological, behavioral and morphological traits, that can impart changing degrees of fitness as a function of the particular abiotic and biotic conditions.

We have rigorous mathematical proof (see Appendices \ref{a1} \& \ref{a2}) that according to the rules governed by this game theoretic model, only those species that possess maximal variability amongst individual organisms have a unique quality:  there is no other species with the same (or lower) mean competitive ability which has positive expected value in competition. In other words, there is no species that can outcompete the maximally variable species.  On the other hand, for any species which does not possess maximal variability amongst individual organisms, in the proof of Theorem 1, we explicitly construct a species with the same (or lower) mean competitive ability which has positive expected value in competition, whereas the original species has negative expected value.  The maximally variable species is the only species which does not have such vulnerability.  Moreover, in Proposition 1, we prove  that such maximally variable species are neutral towards each other in competition for limited resources; they coexist.    

Over time, if species tend toward maximum variability, then they simultaneously tend towards neutral coexistence.  In this sense, we provide a plausible mathematical explanation for two widely documented but poorly understood characteristics of microbes: (1) enormous intra-specific variability and (2) tremendous biodiversity.  


This work is organized as follows. In section 2 we introduce our game theoretic model for microbes.  This sets the stage for our mathematical results which are presented in section 3.  Concluding remarks are offered in section 4.  The complete, rigorous mathematical proofs of the results presented in section 3 are contained in appendix A and appendix B.

\section{Game theory for microbes}
The mathematics behind our approach is non-cooperative game theory.  This is justified by the lack of evidence that  individual microbes communicate in such a way as to create coalitions or to cooperate.  Biofilms are, however, an exception \cite{Flemingetal2016}.  Competition occurs between individuals. The outcome of competition between individuals, no matter how biologically complex, can be reduced to three possibilities:  win, lose, or draw (see Figure 1).  Due to the asexual reproduction of microbes, cumulative success of individuals in competition can be interpreted as population increase, cumulative losses lead to population decrease, and cumulative draws to neither gains nor losses.
\begin{figure}  \label{game-pic} \begin{center} \includegraphics[width=0.9\textwidth]{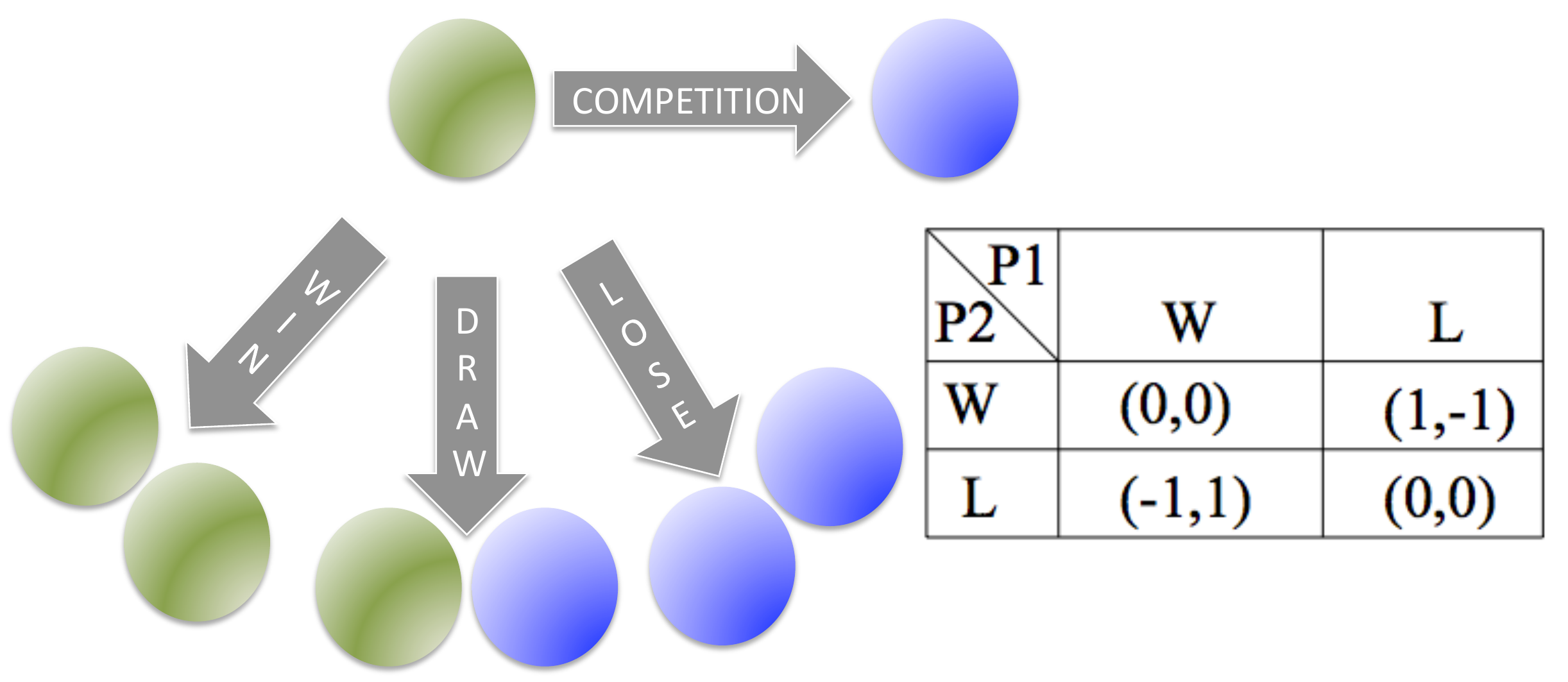} \caption{When two individuals compete, there are three possible outcomes for each individual:  win=self-replicate; lose=die; draw=maintain status quo for each competitor.  The table shows the pay-off function of individual level competition.  A payoff +1 corresponds to win, a payoff -1 corresponds to lose, and a payoff 0 corresponds to a draw.} \end{center}  \end{figure} 

In the game theoretic approach to examine competition on the level of species, we assume that each species is comprised of several individual organisms.  Each individual has a competitive ability, which is not necessarily constant, but may change from one round of competition to the next. The competitive ability (CA) is represented by a number between 0 and 1. We assume that the species as a whole has a mean competitive ability (MCA) which does not exceed 1/2.  The range of both the competitive abilities as well as the mean competitive abilities was chosen arbitrarily and our results are independent of the choice of specific values.  At each round of competition, individual organisms are randomly put in pairs.  The organism with the higher CA doubles whilst the organism with the lower CA perishes; in case of equal CAs both organisms simply persist. The magnitude of difference in CA between competitors is irrelevant. 

A key, and we think novel, feature of our model is that the individual CAs are variable, depending on the distribution of CAs for a given species. Thus, the assembly of the individuals in a species represent the distribution of CAs, and their collective mean CA is equal to the species's MCA. This situation is representative of genetically identical individuals that temporarily differ in competitive ability, due to for example their recent exposure to resources or metabolic demands, the particular state in their life cycle and such. Finally, competition in a dynamic system requires the nature of the competition to vary, for example, competition may be with respect to resource acquisition in one round, and predator defence in another. Thus, individuals that were superior relative to the mean in one round may be inferior within the next round. 

This model applies to any arbitrary number of simultaneously competing species, with arbitrary (not necessarily equal) population sizes.  Each species does not compete in a series of tournaments against all its competitors, but rather, it is competing against all other species simultaneously.  So, our model shows how each species will fare in competition with any arbitrary number of other species.

\subsection{Same-same but different}
For each MCA value, there are numerous ways to assign the individual CAs which have this same MCA value.  We term this ``same-same but different.''  We recall the definition of strategic behavior distribution (SBD) from \cite{jrsi} specified to this setting.  Here, an SBD is defined to be a particular way to assign CAs to individuals such that the MCA is equal to a fixed value. Interestingly, SBDs having the same MCA nonetheless have dramatically different ecological repercussions depending on the particular characteristics of the SBD.  

Largely in order to reduce complexity, it is common to assume that variation among individuals and even species can adequately be represented by a mean or maximal response (e.g. \cite{eppley1972temperature}).  In this case, that would correspond to all individuals being equal and having CA always equal to the MCA. We refer to this degenerate SBD as the invariant SBD.  We shall see in the proof of our main result that it is easily susceptible to invasion by a competing SBD which also has MCA less than or equal to 1/2.

In general, we use here SBDs which are neither purely deterministic nor purely probabilistic. The SBDs are not purely deterministic because the CA of each individual is in general non-constant.  For example, we define the bimodal SBD to select half of the organisms of the species to have CA=1, and the remaining half to have CA=0.  Any odd-man-out gets 0 or 1 with probability 1/2. However, this selection occurs at each round of competition, so an organism which had CA=0 in the first round of competition could have CA=1 in the second round of competition.  Moreover, for an organism which had CA=1 and succeeded in a round of competition and thereby replicates, both it and its clone could have either CA=1 or CA=0 in the next round of competition.  Thus, the model is not deterministic.  On the other hand, the model is also not purely probabilistic in the sense that each individual is not assigned its CA=1 or 0 each with probability 1/2 , with the probabilities of each individual being independent.  If that were the case, then it would be possible, but unlikely, that at a specific round of competition, all individuals had CA=0.  In our model, as long as the number of individuals is at least two, this is impossible.  Thus, the SBDs we use are neither purely deterministic nor purely probabilistic.  
This fluidity of differences among individuals requires the species under consideration to be an asexually reproducing microbial species.  This restricts our results to microbes, because asexual reproduction will quickly regenerate individuals with identical genetic capacity within the SBD of the species, and thus offsetting losses due to an inferior CA.

Below, to determine the success or lack thereof of defined SBDs, we use this game theoretic model to compute the expected value of SBDs in competition.  A positive expected value implies population growth.  A negative expected value implies population decrease.  An expected value of 0 implies stable population, neither increase or decrease.  

\section{Theoretical model and main results} 
The competitive abilities are selected from the values:
$$0, \frac{1}{M}, \frac{2}{M}, \ldots, 1,$$
where the number $M$ above is an integer greater than or equal to $6$.  Here it is convenient to introduce the notation
$$x_j = \frac j M, \quad \textrm{ for } j = 0,1, \ldots, M.$$  
For a species, $A$, comprised of $N_A$ total individual organisms, for $0\leq x_j \leq 1$ we define $n_A(x_j)$ so that 
$$\frac{n_A(x_j)}{N_A}$$
is the probability that an individual organism has competitive ability (CA) equal to $x_j$.  We are interested in the expected value of $A$ in competition with species $B$.  In fact, our model works equally well if we consider $B$ to be all other species which compete with $A$ for limited resources, because we do not require the total number of organisms, $N_A$ and $N_B$, respectively, to be equal.  We do assume that $N_A$ and $N_B$ are both positive.  

\subsection{Expected value in competition} 
We assume that organisms are randomly paired to compete.  Then, we compute the probability that the organisms in species $A$ which have competitive ability equal to $x_i$ compete with an inferior organism by first counting all of the organisms of species $B$ which have lower CA:
$$\sum_{j<i} n_B (x_j).$$
The probability of competing with such an organism is given by dividing by the total number of organisms of species $B$, 
$$\frac{1}{N_B} \sum_{j<i} n_B(x_j).$$
In a similar way, we compute the probability of competing with a superior organism is 
$$\frac{1}{N_B} \sum_{j>i} n_B(x_j).$$
Since we do not assume $N_A$ and $N_B$ are equal, some organisms might not compete at all.  If $N_A<N_B$ then only $N_A$ organisms out of the total $N_B$ organisms in species $B$ actually compete, so the probability that an organism from species $B$ competes is $\frac{N_A}{N_B}$.  In this case, all organisms from species $A$ compete.  If $N_B< N_A$, then only $N_B$ organisms out of the total $N_A$ organisms in species $A$ compete, so the probability that an organism from species $A$ competes is $\frac{N_B}{N_A}$.  In this case, all organisms from species $B$ compete.  So, in general, we define
$$N_c := \min \{ N_A, N_B \} = \textrm{ number of competing organisms.}$$
Then, the probability that an organism from species $A$ actually competes is 
$$\frac{N_c}{N_A},$$
and the probability that an organism from species $B$ actually competes is 
$$\frac{N_c}{N_B}.$$
Hence, the expected value for the competition of the organisms in species $A$ with competitive ability $x_i$ is 
$$\frac{N_c}{N_A} \frac{n_A(x_i)}{N_B} \left( \sum_{j<i} n_B(x_j) - \sum_{j>i} n_B(x_j) \right).$$
The expected value for species $A$ in competition with species $B$ is then calculated by summing the expected value for all competitive abilities:  
\begin{equation} \label{evcalc} E[A,B] = \frac{N_c}{N_A N_B} \sum_{i=0} ^M n_A(x_i) \left( \sum_{j<i} n_B(x_j) - \sum_{j>i} n_B(x_j) \right). \end{equation} 
The competition has a zero-sum dynamic, 
\begin{equation} \label{0sum} E[A,B] + E[B,A] = 0 \implies E[B,A] = - E[A,B]. \end{equation}  
To see why this is true, we use the above considerations to compute:  
$$E[B,A] = \frac{N_c}{N_A N_B} \sum_{i=0} ^M n_B(x_i) \left( \sum_{j<i} n_A(x_j) - \sum_{j>i} n_A(x_j) \right).$$
It is then straightforward to compute that 
$$\sum_{i=0} ^M n_B (x_i) \sum_{j < i} n_A (x_j) = \sum_{k=0} ^M n_A (x_k) \sum_{i > k} n_B (x_i),$$
and similarly
$$\sum_{i=0} ^M n_B (x_i) \sum_{j > i} n_A (x_j) = \sum_{k=0} ^M n_A (x_k) \sum_{i < k} n_B (x_i).$$
Hence 
$$E[B,A] = \frac{N_c}{N_A N_B} \left(  \sum_{k=0} ^M n_A (x_k) \sum_{i > k} n_B (x_i) -    \sum_{k=0} ^M n_A (x_k) \sum_{i < k} n_B (x_i) \right) $$
$$= - \frac{N_c}{N_A N_B} \sum_{k=0} ^M n_A (x_k) \left( \sum_{i < k} n_B (x_i) - \sum_{i > k} n_B (x_i) \right)$$
$$= - E[A,B].$$
This zero-sum dynamic may be interpreted for competition for limited resources.  Furthermore we impose the energy and resource constraint corresponding to the assumption that the MCA does not exceed $1/2$.  The MCA is computed by taking the average of all competitive abilities, so we define 
$$\mca(A) = \frac{1}{N_A} \sum_{i=0} ^M x_i n_A (x_i).$$
Then, the constraint is that for both species 
$$\mca (A) = \frac{1}{N_A} \sum_{i=0} ^M x_i n_A(x_i) \leq \frac{1}{2}, \quad \mca (B) = \frac{1}{N_B} \sum_{i=0} ^M x_i n_B(x_i) \leq \frac{1}{2}.$$

\subsection{The uniform SBD}  
Our first result shows that a certain SBD, known as the \em uniform SBD, \em is neutral in competition to all other SBDs which have the same MCA as that of the uniform SBD.  At the same time, we also show that the uniform SBD has positive expected value in competition with any SBD whose MCA is less than the MCA of the uniform SBD.  
\begin{proposition} Let $A$ be a species with MCA at most $\frac 1 2$.  Let $U$ be the uniform SBD,  which is the unique SBD having $n_A(x_i)$ equal and positive for all $i=0, \ldots, M$.  Then $\mca(U) = \frac 1 2$, and 
$$E[A, U] \leq 0,$$
with equality if and only if the MCA of species $A$ is equal to $\frac 1 2$.  We note that this is independent of the number of organisms in species $A$ and species $U$.   
\end{proposition} 

\subsubsection{Logical argumentation}
To see why this is true, we use the definition of $E[A,U]$ to write out this expected value; see \eqref{eau} in Appendix \ref{a1}.  At first, the expression may look complicated, but using the definition of the uniform SBD, as done in \eqref{defunif} of Appendix \ref{a1}, the expression simplifies.  With these mathematical manipulations and resulting simplifications, the expression for the expected value of species $A$ in competition with the uniform species $U$ becomes 
$$E[A,U] = \frac{N_c M}{M+1} \left( 2 \mca (A) - 1 \right).$$
Above, $N_c$ is the number of competing organisms, which we recall is the smaller of $N_A$ and $N_U$.  The number $M$ is how we have defined the competitive values, as the numbers $0, \frac 1 M, \frac 2 M$, and up to $1 = \frac M M$.  In the parentheses we see $2 \mca(A)$.  By the constraint on the MCA, we know that $\mca (A) \leq \frac 1 2$, so $2 \mca(A) \leq 1$.  Consequently, 
$$(2 \mca (A) - 1) \leq 0, \textrm{ so we also see that } E[A,U] = \frac{N_c M}{M+1} \left( 2 \mca (A) - 1 \right) \leq 0.$$
This shows that $E[A,U] \leq 0$.  Next, if $\mca(A) < \frac 1 2$, then 
$$(2 \mca (A) - 1) < 0, \textrm{ and consequently } E[A,U] = \frac{N_c M}{M+1} \left( 2 \mca (A) - 1 \right)  < 0.$$
If instead, $\mca (A) = \frac 1 2$, then $2 \mca(A) - 1 = 0$, and consequently the expression for $E[A,U]$ is zero.

\subsection{Non-exploitable strategies} 
Here, we introduce a new definition which is a generalization of an evolutionary stable strategy.  Recall that an evolutionary stable strategy (ESS), $A$, must satisfy for all other strategies, $B$, 
$$E[A,A] > E[B,A] \textrm{ or } E[A,A] = E[B,A] \textrm{ and } E[A,B] > E[B,B].$$

\begin{definition} \label{nes} A strategy, $A$, is a \em non-exploitable strategy, \em abbreviated NES, if and only if 
$$E[A,B] \geq E[B,A] \quad \textrm{ for all strategies, $B$.}$$
\end{definition} 
\begin{lemma} Every ESS is also an NES in our setting.
\end{lemma}

\subsubsection{Logical argumentation}
We start by assuming that $A$ is an ESS.  By the definition of an ESS, $E[A,A] \geq E[B,A]$ for all strategies, $B$.  Due to the zero-sum dynamic,  $E[A,A] = 0$.  So, this means that 
$$0 \geq E[B,A] \textrm{ for all strategies, $B$.}$$  
By the zero-sum dynamic, 
$$E[A,B] + E[B,A] = 0,$$
so 
$$E[A,B] = - E[B,A].$$
Since $E[B,A] \leq 0$, it follows that $- E[B,A] \geq 0$.  Therefore, 
$$E[A,B] = - E[B,A] \geq 0.$$
This holds for all strategies, $B$.  We use the zero-sum dynamic once more to see that 
$$E[A,B] = - E[B,A] \geq 0 \geq E[B,A] \textrm{ for all strategies, $B$.}$$  
We therefore see that the definition for $A$ to be an NES holds true.  Hence, we see that every ESS is also an NES, but we shall see below that not every NES is an ESS.  

\subsection{The unique NES}  
Our main theorem proves that (1) there is no ESS in our setting, but (2) the uniform SBD is the unique NES. 

\begin{theorem} \label{okyvliganda} Let $A$ be a species comprised of $N\geq 4$ individual organisms and which has MCA less than or equal to $1/2$.  Assume that $M \geq 6$.  Then, there are two \em mutually exclusive \em possibilities:
\begin{enumerate} 
\item $A$ is the uniform SBD.  
\item There is an SBD for a species $B$ comprised of at most $N$ individuals which has MCA less than or equal to $1/2$ and which has positive expected value in competition with species $A$.   
\end{enumerate} 
Therefore, we conclude: 
\begin{enumerate} 
\item There is no SBD for species with at least $N \geq 4$ individual organisms and which has MCA less than or equal to $1/2$ which is an ESS.
\item The unique NES is the uniform SBD.
\end{enumerate} 
\end{theorem} 

\subsection{Strategy of the proof}  
We have already demonstrated for the uniform SBD, $U$, that $E[B,U] \leq 0$ for any SBD, $B$, which has MCA less than or equal to $\frac1 2$.  So, the task at hand is to understand species, $A$, which are \em not uniform.  \em   We therefore begin by assuming that $A$ is \em not \em the uniform SBD.  We then proceed with several straightforward steps.  

\subsubsection{First simplification}  We begin by making a simplifying observation.  If $\mca(A) < \frac 1 2$, then we have shown that the uniform species has positive expected value in competition with species $A$.  So, in that case, the theorem is proven.  To complete the proof, we can henceforth assume that $\mca(A) = \frac 1 2$.  

\subsubsection{Pairs of CAs}  The next idea is to look at the number of organisms at pairs of competitive abilities, starting with the lowest $CA = 0$ together with the highest $CA = 1$.  We look at $n_A (x_0)$ and $n_A (x_M)$.  There are three possibilities:  either $n_A(x_0) > n_A (x_M)$ or $n_A (x_0) < n_A (x_M)$, or $n_A (x_0) = n_A (x_1)$.   
\begin{enumerate} 
\item In the first case, if $n_A (x_0) > n_A (x_M)$, we define species $B$ to have $n_B (x_0) = N/2$ and $n_B (x_M) = N/2$, where $N$ is the total number of organisms as in the theorem.  Then, we compute the expected value of $B$ in competition versus $A$ in \eqref{evbi} of \S \ref{pocas}, 
$$E[B,A] = \frac{n_A (0) - n_A (1)}{2} > 0.$$
So, in this case we have found a species $B$ which has MCA at most $\frac 1 2$ (in fact equal to $\frac 1 2$), and the same number of organisms as species $A$, but which has positive expected value in competition versus species $A$.  
\item In the second case, $n_A (x_0) < n_A (x_M)$.  Here we define $B$ with
$$ n_B (x_j) := \begin{cases} 0 & x_j = 0 \\ 0 & x_j = 1 \\ \frac{N}{M-1} & \textrm{ otherwise.} \end{cases} $$
We compute in \S \ref{pocas} that 
$$E[B,A] = \frac{1}{M-1} \left( n_A(x_M) - n_A(x_0) \right) > 0,$$
since $n_A (x_M) > n_A (x_0)$.  
\item In the third case, $n_A (x_0) = n_A (x_M)$.  In this case we don't do anything, but instead proceed to look at $n_A (x_1)$ and $n_A (x_{M-1})$.  
\end{enumerate}

\subsubsection{Pair-by-pair}  We look next at $n_A (x_1)$ and $n_A (x_{M-1})$.  There are the same three cases as above.  In the first and second cases, we can define a species $B$ which has $\mca(B) \leq \frac 1 2$, and the total number of organisms of species $B$ does not exceed $N$.  
If $n_A\left( \frac 1 M \right) > n_A \left( 1 - \frac 1 M \right)$ we define the species $B$ to have $n_B (x_1) = \frac N 2$, and $n_B (x_{M-1}) = \frac N 2$.  Then, we compute in \S \ref{pbp} that the expected value of $B$ in competition versus $A$ is 
$$E[B, A] = \frac{ n_A \left( \frac 1 M \right) - n_A \left( 1 - \frac 1 M \right)}{2} > 0.$$
On the other hand, if $n_A(x_1) < n_A(x_{M-1})$, then we define 
$$n_B (x_j) = \begin{cases} \frac{N}{M-3} & \frac 1 M < j < 1- \frac 1 M \\ 0 & \textrm{ otherwise } \end{cases}.$$
In \S \ref{pbp} we show that the number of organisms in species $B$ is $N$, and the MCA of species $B$ is $\frac 1 2$.  We compute there that 
$$E[B,A] =  \left( \frac{M-2}{M-3} \right) \left( n_A (x_{M-1}) - n_A (x_1) \right) - \left( n_A (x_{M-1}) - n_A (x_1) \right)   > 0,$$
In the last case, $n_A (x_1) = n_A(x_{M-1})$, we do not do anything, just look at the next pair, looking at $n_A (x_2)$ and $n_A (x_{M-2})$.  In this way, we continue pair-by-pair.

\subsubsection{Second simplification}  There are now two mutually exclusive possibilities:  either at some step we found a pair which is lopsided, that is $n_A (x_k) \neq n_A (x_{M-k})$ for some $k=0,1,\ldots$.  Once we reach such a pair, we show in \S \ref{sss} how to construct the species $B$ which has $E[B,A] > 0$.  So, in all such cases, the theorem has been proven.  It therefore only remains to consider the case when 
$$n_A (x_k) = n_A (x_{M-k})$$
for all $k$.  

\subsubsection{Exploiting non-uniformity} We are now reduced to the case where $n_A (x_k) = n_A (x_{M-k})$ holds for all $k$.  That means that the SBD is symmetric around the competitive ability $\frac 1 2$.  However, we assumed that $A$ is \em not-uniform.  \em  This means that the numbers $n_A(x_k)$ cannot all be identical.  So, some of the $n_A (x_k)$ are larger, and some are smaller.  It is precisely this lack of uniformity which allows us to find a weakness and exploit it with species $B$.  The details of this process are contained in \S \ref{enu}.

\section{Concluding remarks}
Evolution, or survival of the fittest, in the case of microbes must be approached from a different perspective in comparison to macro-organisms.  Rather than assessing the fitness of individual microbes and applying evolution to individuals interpreted as representing a species, we propose that it is more fitting to consider cohorts of individuals that collectively but not individually represent the entire species.  Individual microbes do not always behave in a way that maximizes fitness, nor does loss of an individual necessarily result in the loss of the genotype, because  asexual reproduction yields multiple clones and can rapidly increase the abundance of specific clones or replace lost individuals representing a particular clone. Thus, loss of individuals can have no consequence for the species as a whole, suggesting a very different risk distribution, where the species as a whole may nonetheless survive and remain variable with respect to their trait distribution.  Thus, we analyze evolutionary fitness of species by interpolating between the individual-level competitions and the species-level growth or attrition.  

As we are assuming competition for limited resources, if one species has a positive expected value, implying population growth, the other species has a negative expected value, implying population decrease.  Presuming that species evolve so as to minimize vulnerability to invasion and attrition by other species, rather than to maximize fitness, then all species of microbes which survive would evolve to the uniform SBD.  Obviously, the rate of change in the SBD will be influenced by the environmental conditions, the presence of co-occurring species and to what degree SBDs are similar or overlap with co-occurring species (e.g. character displacement). Irrespective of any external factors, the uniform SBD is the unique SBD which contains the maximal variability amongst its individuals.  Moreover, it is the unique SBD which is not vulnerable to replacement by species with other SBDs with the same (or lower) MCA.  There are two mutually exclusive options:  evolve towards the uniform SBD or risk vulnerability to invasion.  

Interestingly, the uniform SBD is also neutral in competition with all other SBDs which have the same mean competitive ability.  Thus, our second result shows that the uniform SBD coexists with all other SBDs which have the same MCA.  Hence, it predicts that microbes which evolve to the uniform SBD simultaneously evolve to co-exist.
We therefore have discovered a plausible mathematical explanation for two prevalent characteristics of microbes which until now have been inexplicable in the context of competition theory: (1) maximal variability of traits amongst individuals of a single species and (2) co-existence of virtually limitless different species of microbes. For multi-cellular organisms, intra-specific variability has been suggested to be an important characteristic in the management of biodiversity \cite{desRoches2018}. Our results imply that a quantification of the trait distributions of microbial organisms is essential to understand processes that drive ecosystem functioning, including ocean ecosystems and the global carbon cycle.

\begin{acknowledgements}
Susanne Menden-Deuer was supported through the National Aeronautics and Space Administration (award \#NNX15AL2G).  Julie Rowlett is grateful to Ksenia Fedosova, Philip Gerlee, and Jeffrey Steif for stimulating discussions and insightful comments.  Many thanks to Medet Nursultanov for carefully checking the proofs and for suggesting the SBDs defined in \eqref{middle} and \eqref{n-1}.  We are also grateful to Edvin Wedlin for suggesting the term ``non-exploitable strategy,'' NES.  Constructive feedback by three anonymous reviewers has improved a prior version of this manuscript.  
\end{acknowledgements}



\begin{appendix}

\section{Proof of Proposition 1} \label{a1} 
We compute the expected value, 
\begin{equation} \label{eau} E[A, U] = \frac{N_c}{N_A N_u} \sum_{i=0} ^M n_A (x_i)  \left( \sum_{j < i} n_u (x_j) - \sum_{j > i} n_u (x_j) \right), \end{equation} 
where above $N_u$ denotes the total number of uniform organisms, and $n_u (x_j)$ denotes those with competitive ability equal to $x_j$.  Similarly, we use $N_A$ to denote the total number of organisms of species A, and 
$$N_c = \min\{N_u, N_A\}$$
is the number of competing organisms.  By definition of the uniform SBD, 
\begin{equation} \label{defunif} n_U (x_j) = \frac{N_u}{M+1} \forall j \implies \sum_{j < i} n_u (x_j) = \sum_{j=0} ^{i-1} \frac{N_u}{M+1} = \frac{i N_u}{M+1}. \end{equation} 
Similarly, 
$$\sum_{j > i} n_u (x_j) = \sum_{j=i+1} ^M \frac{N_u}{M+1} = \frac{ (M-i) N_u}{M+1}.$$
Hence, 
$$E[A, U] = \frac{N_c}{N_A N_u} \sum_{i=0} ^M n_A(x_i) \left( \frac{i N_u}{M+1} - \frac{(M-i) N_u}{M+1} \right)$$ 
$$= \frac{N_c}{N_A (M+1)} \sum_{i=0} ^M n_A (x_i) \left( 2i - M \right) = \frac{N_c M}{M+1} \left( 2 \mca (A) - 1 \right) \leq 0.$$
The last inequality follows because $\mca(A) \leq \frac 1 2$.  So, we see that equality holds if and only if $\mca(A) = \frac 1 2$.  
\qed

\section{Proof of Theorem 1} \label{a2} 
Let $A$ be a species comprised of $N$ individual organisms.  Let $x_k = \frac{k}{M}$ for $0\leq k \leq M$, and $n_A(x_k)$ be the number of organisms of species $A$ with competitive ability $x_k$.  
\subsection{First simplification} \label{simp1} 
If the MCA of $A$ is less than $\frac 1 2$, then the uniform SBD has MCA equal to 1/2, and by Proposition 1, 
$$E[U,A] > 0.$$
So, we may henceforth assume that the MCA of species $A$ is in fact equal to $1/2$.   We shall prove the result by an induction argument.  In this argument, we will be considering pairs of competitive abilities, $(x_k, x_{M-k})$.  To begin, we consider the SBD in which species $A$ has half of its organisms with competitive ability equal to 0, and the other half with competitive ability equal to 1.  In case $N$ is odd, any ``odd man out'' has competitive ability 0 or 1 each with probability 1/2.   We call this the bimodal SBD.  

In this case, we define species $B$ to have 
$$n_B (x_k) = \begin{cases} 1 & k=M \\ N-1 & k=1 \\ 0 & k \neq M \textrm{ and } k \neq 1. \end{cases}$$
Then, we compute that the MCA of this species is 
$$\frac{1 + (N-1) \frac 1 M}{N} = \frac 1 N + \frac 1 M - \frac 1 {NM} \leq \frac 1 N + \frac 1 M \leq \frac 1 4 + \frac 1 4 \leq \frac 1 2.$$
We have used above the assumption that $M$ and $N$ are both greater than or equal to $4$.  We compute the expected value 
$$E[B, A] = \frac 1 N \frac N 2 - \frac{N-1}{N} \left( \frac N 2 - \frac N 2\right) = \frac 1 2 > 0.$$

Henceforth, we may assume that the SBD of species $A$ is not the bimodal SBD.  By this assumption $n_A(0)$ and $n_A(1)$ cannot both be equal to $N/2$.  It will now be convenient to introduce the notation 
\begin{equation} \label{knotation} n_k := n_A \left( \frac 1 2 + \frac k M \right), \quad k \in \Z. \end{equation} 
Moreover, we will henceforth assume that $M$ is even.  Were $M$ not even, we would simply make the partition twice as fine.  Then, we use the constraint that the MCA is equal to $\frac 1 2$ to compute:
$$ \frac{ \ldots + \left( \frac 1 2 - \frac k M \right) n_{-k} + \ldots + \left( \frac 1 2 - \frac 1 M \right) n_{-1} + \frac 1 2 n_0 + \left( \frac 1 2 + \frac 1 M \right) n_{1} + \ldots + \left( \frac 1 2 + \frac k M\right) n_{k} + \ldots }{\ldots + n_{-k} + \ldots + n_{-1} + n_0 + n_1 + \ldots + n_k + \ldots} = \frac 1 2.$$
Re-arranging, 
$$\sum_{k=1} ^{M/2} \left( \frac 1 2 - \frac k M \right) n_{-k} + \frac 1 2 n_0 + \sum_{k=1} ^{M/2} \left( \frac 1 2 + \frac k M \right)n_k = \frac{\sum_{k=1} ^{M/2} n_{-k} + n_0 + \sum_{k=1} ^{M/2} n_k }{2}.$$
Grouping the terms on the left gives 
$$\frac 1 2 \left( \sum_{k=1} ^{M/2} n_{-k} + n_0 + \sum_{k=1} ^{M/2} n_k \right) + \sum_{k=1} ^{M/2} \frac k M \left(n_k - n_{-k} \right) = \frac{\sum_{k=1} ^{M/2} n_{-k} + n_0 + \sum_{k=1} ^{M/2} n_k }{2}.$$
Hence, we obtain the rather useful identity that 
\begin{equation} \label{keq} \sum_{k=1} ^{M/2} \frac k M \left(n_k - n_{-k} \right)  = 0 \implies \sum_{k=1} ^{M/2} k \left(n_k - n_{-k} \right) = 0. \end{equation} 

\subsection{Pairs of CAs} \label{pocas}
We proceed to look at $n_A (0)$ and $n_A (1)$.   If 
\begin{equation} \label{usebimodal} n_A (0) > n_A (1), \quad \textrm{ let $B$ be the bimodal SBD.} \end{equation}
If this is the case, we compute the expected value of $B$ in competition against $A$ is 
\begin{equation} \label{evbi} E[B,A] = \frac{N}{2N} \left( N - n_A (1) \right) - \frac{N}{2N} \left( N - n_A (0) \right) = \frac{n_A (0) - n_A (1)}{2} > 0. \end{equation} 
Hence, the bimodal species satisfies the statement of the theorem in competition against species $A$.  

If on the other hand we have 
\begin{equation} \label{usewindow} n_A(0) < n_A (1), \quad \textrm{ define $B$ with } n_B (x_j) := \begin{cases} 0 & x_j = 0 \\ 0 & x_j = 1 \\ \frac{N}{M-1} & \textrm{ otherwise.} \end{cases} \end{equation} 
We then compute the expected value of $A$ in competition versus $B$.  Since both species have $N$ organisms total, $E[A,B] = - E[B,A]$, so this computation actually yields both values.  The $n_A(1)$ individuals with CA equal to 1 always win, whereas the $n_A(0)$ individuals with CA equal to 0 always lose.  Hence, they contribute 
$$n_A (1) - n_A(0) \textrm{ to the expected value of $A$ versus $B$.}$$
We now compute the expected value from the other individuals.  For this we recall the notation \eqref{knotation}.  For an individual of species $A$ with CA equal to $\frac 1 2 + \frac k M$, we count that there are 
$$\left( \frac M 2 - 1 - k\right) \frac{N}{M-1}$$
individuals of species $B$ which have strictly higher competitive abilities.  We similarly count that there are 
$$\left( \frac M 2 - 1 + k \right) \frac{N}{M-1}$$
individuals of species $B$ which have strictly lower competitive abilities.  Hence, the expected value from the individuals of species $A$ with CA equal to $\frac 1 2 + \frac k M$ is 
$$n_k \frac 1 N \left( \left( \frac M 2 - 1 + k \right) \frac{N}{M-1} - \left( \frac M 2 - 1 - k\right) \frac{N}{M-1}\right) = n_k \frac{2k}{M-1}.$$
By the symmetry in the definition of species $B$, for an individual of species $A$ with CA equal to $\frac 1 2 - \frac k M$, these numbers are switched, so we compute the expected value is 
$$n_{-k} \frac 1 N \left( \left( \frac M 2 - 1 - k\right) \frac{N}{M-1} -  \left( \frac M 2 - 1 + k \right) \frac{N}{M-1} \right) = n_{-k} \frac{-2k}{M-1}.$$
We therefore compute the total expected value, 
$$E[A,B] = n_A (1) - n_A(0) + \sum_{k=1} ^{\frac M 2 - 1} \frac{2k}{M-1} \left(n_k - n_{-k} \right) $$
$$= n_A (1) - n_A(0) + \frac{2}{M-1}  \sum_{k=1} ^{\frac M 2 - 1} k\left(n_k - n_{-k} \right)  .$$
Now, we see that \eqref{keq} is quite useful: 
$$\sum_{k=1} ^{M/2} k \left(n_k - n_{-k} \right) = 0 \implies - \frac M 2 \left(n_{M/2} - n_{-M/2} \right) = \sum_{k=1} ^{\frac M 2 - 1} k (n_k - n_{-k} ).$$
Recall that by definition \eqref{knotation}
$$n_A(1) = n_{M/2}, \quad n_A (0) = n_{-M/2}.$$
Hence, 
$$E[A,B] = n_A (1) - n_A (0) - \frac{2M}{2(M-1)} \left( n_A(1) - n_A(0)\right) = \frac{(M-1 - M)  \left( n_A(1) - n_A(0)\right)}{M-1}$$
$$= -\frac{1}{M-1} \left( n_A(1) - n_A(0) \right) < 0, \quad \textrm{ since } n_A(1) > n_A(0).$$

So, in case $n_A(0)$ and $n_A(1)$ are unequal, or in case $A$ is the bimodal SBD, we have proven the theorem.  

\subsection{Pair-by-pair} \label{pbp} 
We now proceed inductively.  Assume that $n_A(0)$ and $n_A(1)$ are identical.  We now consider 
$$n_A\left( \frac 1 M \right) \textrm{ and } n_A \left( 1 - \frac 1 M \right).$$
If these are unequal, in the same way as above, we can construct a species, $B$, which has higher expected value in competition.  In case $n_A\left( \frac 1 M \right) > n_A \left( 1 - \frac 1 M \right)$ we define the species $B$ to have its organisms equally distributed at $\frac 1 M$ and $1-\frac 1 M$.  By the same calculation as above, we compute the expected value 
$$E[B, A] = \frac{ n_A \left( \frac 1 M \right) - n_A \left( 1 - \frac 1 M \right)}{2} > 0.$$
On the other hand, if 
\begin{equation} \label{inds1} n_A\left( \frac 1 M \right) < n_A \left( 1- \frac 1 M \right), \end{equation} 
we define 
$$n_B (x_j) = \begin{cases} \frac{N}{M-3} & \frac 1 M < j < 1- \frac 1 M \\ 0 & \textrm{ otherwise } \end{cases}.$$
Then we compute that the number of organisms in species $B$ is 
$$\sum_{j=0} ^M n_B (x_j) = \sum_{j=2} ^{M-2} n_B (x_j) = \frac{N (M-3)}{M-3} = N.$$
Moreover, so defined the MCA of species $B$ is $\frac 1 2$.  In species, $A$, the $n_A (1)$ organisms with CA equal to one always win, whereas the $n_A (0)$ organisms with CA equal to 0 always lose.  Hence their cumulative contribution to $E[A,B]$ is zero.  We therefore compute 
$$E[A,B] = n_A (x_{M-1}) - n_A (x_1) + \sum_{k=1} ^{\frac M 2 - 2} \frac{2k}{M-3} (n_k - n_{-k}).$$
Since $n_A(x_1) - n_A(x_0) = 0$, 
$$ \sum_{k=1} ^{\frac M 2 - 2} \frac{2k}{M-3} (n_k - n_{-k}) = \frac{2}{M-3} \sum_{k=1} ^{\frac M 2} k (n_k - n_{-k}) - \frac{2}{M-3} \left( \frac M 2 - 1 \right) \left( n_A (x_{M-1}) - n_A (x_1) \right).$$
By \eqref{keq}, this is 
$$- \frac{2}{M-3} \left( \frac M 2 - 1 \right) \left( n_A (x_{M-1}) - n_A (x_1) \right) = - \left( \frac{M-2}{M-3} \right) \left( n_A (x_{M-1}) - n_A (x_1) \right).$$
Hence, we have computed that
$$E[A,B] = \left( n_A (x_{M-1}) - n_A (x_1) \right)  - \left( \frac{M-2}{M-3} \right) \left( n_A (x_{M-1}) - n_A (x_1) \right) < 0,$$
where the final inequality follows from \eqref{inds1} together with the observation that $\frac{M-2}{M-3} > 1$.  

We continue by induction.  We assume that $n_A (x_j) = n_A (x_{M-j})$ for all $j=0, \ldots, k-1$.  Then, if $n_A (x_k) > n_A (x_{M-k})$, we define 
$$n_B (x_j) = \begin{cases} \frac N 2 & j=k \\ \frac N 2 & j=M-k \\ 0 & \textrm{ otherwise } \end{cases} .$$
Similar to our previous calculations, 
$$E[B,A] = \frac{n_A (x_k) - n_A (x_{M-k})}{2} > 0.$$ 
If on the other hand we have $n_A (x_k) < n_A (x_{M-k})$, we define 
$$n_B (x_j) = \begin{cases} \frac{N}{M-1-2k} & k<j<M-k \\ 0 & \textrm{ otherwise } \end{cases}.$$ 
We compute that species B has $N$ organisms and its MCA is $\frac 1 2$.  Similar to our previous calculations, we compute 
$$E[A,B] = n_A (x_{M-k}) - n_A (x_k) + \sum_{j=1} ^{\frac M 2} \frac{2j}{M-1-2k} (n_j - n_{-j}) - \frac{2 (M/2 - k)}{M-1-2k} \left( n_A (x_{M-k}) - n_A (x_k) \right).$$
Above we have used the assumption that $n_A (x_j) = n_A (x_{M-j})$ for all $j=0, \ldots, k-1$.  By \eqref{keq}, 
$$E[A,B] = \left( n_A (x_{M-k}) - n_A (x_k)  \right) - \frac{M-2k}{M-2k-1}  \left( n_A (x_{M-k}) - n_A (x_k) \right) < 0,$$
where the last inequality follows since $n_A (x_{M-k}) > n_A (x_k)$ and $\frac{M-2k}{M-2k-1} > 1$.  

\subsection{Second simplification} \label{sss} 
We therefore see that at the $k^{th}$ step, if $n_A (x_k) \neq n_A (x_{M-k})$ we can construct a species B such that $E[B,A] > 0$.  So, we can now reduce to the case in which 
$$n_A (x_k) = n_A (x_{M-k}), \quad k=0, 1, \ldots, \frac M 2 - 1.$$ 
In other words, the SBD of species A is symmetric about the competitive value $\frac 1 2$.  In the notation \eqref{knotation}, 
$$n_k = n_{-k}, \quad k = 1, \ldots, \frac M 2.$$

\subsection{Exploiting non-uniformity} \label{enu} 
At this point, we assume that the species A does not have the uniform distribution.  This means that the values $\{n_k \}_{k=0} ^{M/2}$ are not all equal.  So, there is at least one $k$, or perhaps several, for which $n_k$ (and by symmetry, also $n_{-k}$) has the maximum value.  These will be used, together with the assumption that the distribution is not uniform to construct the species, $B$.  First, assume 
\begin{equation} \label{0max} 0 \in \{ j : n_A (x_j) \geq n_A (x_k) \forall k \}. \end{equation} 
This means that $n_A (0)$ and $n_A (1)$ are maximal.  For the uniform distribution, denote by $n_U = n_U (x_k)$ the number of organisms of competitive ability $x_k$, noting that it is the same for all $k$, hence 
$$\sum_{k=2} ^{M-1} n_U (x_k) = (M-2) n_U.$$
However, since species A is \em not \em uniform, and $n_A(0) = n_A (1)$ is maximal, 
\begin{equation} \label{useful} \sum_{k=2} ^{M-1} n_A (x_k) \leq (M-2) n_A (0) = (M-2) n_A (1). \end{equation} 
This will be important in our calculation of the expected value for species, $B$, defined below.  We set 
$$n_B (x_j) = \begin{cases} \left( \frac 1 2 + \eps_0 \right)N & x_j = \frac 1 M \\ \left( \frac 1 2 - \eps_1 \right)N & x_j = 1 \\
0 & x_j \not\in \left\{ \frac 1 M, 1 \right\} .\end{cases}$$
We first verify that the species, B, has at most $N$ organisms:
$$\left( \frac 1 2 + \eps_0 \right)N  + \left( \frac 1 2 - \eps_1 \right)N \leq N \iff \eps_0 - \eps_1 \leq 0 \iff \eps_0 \leq \eps_1.$$
Next, we verify that the MCA of species, B, is at most $\frac 1 2$, 
$$\left( \frac 1 2 + \eps_0 \right) \frac 1 M + \left( \frac 1 2 - \eps_1 \right) \leq \frac 1 2$$
$$\iff \frac{1 + 2\eps_0}{2M} - \eps_1 \leq 0 \iff \frac{1+2\eps_0}{2M} \leq \eps_1.$$
So, we define 
\begin{equation} \label{eps1} \eps_1 := \frac{1+2\eps_0}{2M}. \end{equation} 
Then, species B has MCA precisely equal to $\frac 1 2$.  We must show that this is consistent with the necessary condition that $\eps_0 \leq \eps_1$:
$$\eps_0 \leq \eps_1 = \frac{1+2\eps_0}{2 M} \iff \eps_0 \left( 1- \frac 1 M \right) \leq \frac 1 {2M} \iff \eps_0 \leq \frac{1}{2(M-1)}.$$
So, we will choose $\eps_0 > 0$ sufficiently small so that $\eps_0 \leq \frac{1}{2(M-1)}$.  Since $M$ is fixed, this is certainly possible.  

Since the total number of organisms of species $A$ is $N=N_A$, whereas the total number of organisms of species $B$ is 
$$N_B = \left( \frac 1 2 + \eps_0 \right) N + \left( \frac 1 2 - \eps_1 \right) N \leq N,$$
we see that 
$$N_c = N_B \implies \frac{N_c}{N N_B} = \frac{1}{N}.$$
We therefore calculate the expected value
$$E[B, A] = \left( \frac 1 2 + \eps_0 \right) \left( n_A (0) - \sum_{k=2} ^M n_A (x_k) \right) + \left( \frac 1 2 - \eps_1 \right) \sum_{k=0} ^{M-1} n_A (x_k).$$ 
By symmetry, $n_A (0) = n_A (1) = n_A (x_M)$, so we have 
$$E[B, A] = \left( \frac 1 2 + \eps_0 \right) \left( - \sum_{k=2} ^{M-1} n_A (x_k) \right) + \left( \frac 1 2 - \eps_1 \right) \left( n_A (0) + n_A \left( \frac 1 M \right) + \sum_{k=2} ^{M-1} n_A (x_k) \right)$$
$$= \left( \frac 1 2 - \eps_1 - \frac 1 2 - \eps_0 \right) \sum_{k=2} ^{M-1} n_A (x_k) +  \left( \frac 1 2 - \eps_1 \right) \left( n_A (0) + n_A \left( \frac 1 M \right) \right)$$
$$= (-\eps_0 - \eps_1) \sum_{k=2} ^{M-1} n_A (x_k) +  \left( \frac 1 2 - \eps_1 \right) \left( n_A (0) + n_A \left( \frac 1 M \right) \right).$$
We recall the definition of $\eps_1$ in \eqref{eps1} to write 
$$- \eps_0  -\eps_1 = - \eps_0 - \left( \frac{1+2\eps_0}{2M} \right) = \frac{-2\eps_0 - 2M\eps_0 - 1}{2M}, \quad \frac 1 2 - \eps_1 = \frac{M-2\eps_0 - 1}{2M}.$$
Hence 
$$E[B, A] = \left(  \frac{-2\eps_0 - 2M\eps_0 - 1}{2M} \right) \sum_{k=2} ^{M-1} n_A (x_k)  +  \frac{M-2\eps_0 - 1}{2M} \left( n_A (0) + n_A \left( \frac 1 M \right) \right).$$
Thus 
$$2M \left( E[B,A]\right) = \left( -2\eps_0 - 2M\eps_0 - 1\right) \sum_{k=2} ^{M-1} n_A (x_k)  + \left( M - 1 - 2 \eps_0\right) \left( n_A (0) + n_A \left( \frac 1 M \right) \right).$$
Clearly, 
$$E[B, A] > 0 \iff 2M \left( E[B,A] \right) > 0,$$
since $M>0$.  So, we must show that 
$$\left( 2\eps_0 + 2M\eps_0 + 1\right) \sum_{k=2} ^{M-1} n_A (x_k) < \left( M - 1 - 2 \eps_0\right) \left( n_A (0) + n_A \left( \frac 1 M \right) \right),$$
which is equivalent to  
$$\eps_0 \left(  \left(2 + 2M \right) \sum_{k=2} ^{M-1} n_A (x_k) + 2 \left( n_A (0) + n_A \left( \frac 1 M \right) \right)\right) + \sum_{k=2} ^{M-1} n_A (x_k) $$
$$< (M-1) \left( n_A (0) + n_A \left( \frac 1 M \right) \right).$$
At this point we recall the inequality \eqref{useful}, which implies 
$$(M-1) n_A (0) = (M-2)n_A (0) + n_A (0) \geq \sum_{k=2} ^{M-1} n_A (x_k) + n_A (0).$$
Hence, 
$$(M-1) \left( n_A (0) + n_A \left( \frac 1 M \right) \right) \geq \sum_{k=2} ^{M-1} n_A (x_k) + n_A (0).$$
So, it is sufficient to demonstrate the inequality 
$$\eps_0 \left(  \left(2 + 2M \right) \sum_{k=2} ^{M-1} n_A (x_k) + 2 \left( n_A (0) + n_A \left( \frac 1 M \right) \right)\right) + \sum_{k=2} ^{M-1} n_A (x_k) $$
$$<  \sum_{k=2} ^{M-1} n_A (x_k) + n_A (0)$$
$$\iff \eps_0 \left(  \left(2 + 2M \right) \sum_{k=2} ^{M-1} n_A (x_k) + 2 \left( n_A (0) + n_A \left( \frac 1 M \right) \right)\right) < n_A (0).$$
Since $n_A (0)$ is fixed and positive, we can achieve this inequality by simply taking $\eps_0>0$ sufficiently small.  Specifically, we demand that 
$$0<\eps_0 < \frac{n_A (0)}{ \left(  \left(2 + 2M \right) \sum_{k=2} ^{M-1} n_A (x_k) + 2 \left( n_A (0) + n_A \left( \frac 1 M \right) \right)\right)}.$$

This argument is for the case that the SBD of A is symmetric about $\frac 1 2$, is not the uniform SBD, and we have \eqref{0max}.  For the general case in which the SBD of A is symmetric about $\frac 1 2$ and is not the uniform SBD, let
\begin{equation} \label{jmax} j = \min \{ k : n_A (x_k) \geq n_A (x_l) \forall l \}. \end{equation} 
First, we assume that $j < \frac M 2 - 1$.  Then, we define the SBD of B quite similarly as in the case of \eqref{0max}.  Specifically, we let 
$$n_B (x_k) = \begin{cases} \left( \frac 1 2 + \eps_0 \right)N & x_k = \frac {j+1} M \\ \left( \frac 1 2 - \eps_1 \right)N & x_k= 1-\frac j M \\
0 & \textrm{otherwise} .\end{cases}$$
We similarly require 
$$\eps_0 \leq \eps_1$$
to ensure that the species B has at most N organisms.  To guarantee that the MCA is $\frac 1 2$ we compute 
$$\frac {j+1} M \left( \frac 1 2 + \eps_0 \right) +  \left( \frac 1 2 - \eps_1 \right) \left( 1 - \frac j M \right) = \frac 1 2$$
$$\iff \frac {j+1} M \left( \frac 1 2 + \eps_0 \right) - \frac{j}{2M} - \eps_1 \left( \frac{M-j}{M} \right) = 0$$
$$\iff \frac{j+1+2\eps_0 + 2j\eps_0 - j - 2M\eps_1 + 2j \eps_1}{2M} = 0$$
\begin{equation} \label{eps1j} \iff 1+(2+2j)\eps_0 =  2M \eps_1 - 2j \eps_1 \iff \frac{1+(2 + 2j) \eps_0}{2M - 2j} = \eps_1.\end{equation} 
We check that this is consistent with the requirement that $\eps_0 \leq \eps_1$:
$$\eps_0 \leq \frac{1+(2+2j) \eps_0}{2M - 2j} \iff \eps_0 (2M-2j) \leq 1+(2 +2j)\eps_0 $$
$$\iff \eps_0 (2M - 4j - 2) \leq 1 \iff \eps_0 \leq \frac{1}{2(M -2j - 1)}.$$
We note that this is the same as \eqref{eps1} in case $j=0$.  Above, we see that it is important that we have made the assumption that $j<\frac M 2 - 1$.  This guarantees that 
$$M-2j-1 > M - (M-2) - 1 \geq 1 > 0.$$
Since $j \geq 0$ and $M \geq 4$, it is always possible to choose $\eps_0$ to satisfy these conditions, and to define $\eps_1$ as in \eqref{eps1j}.  So, the species B has at most $N$ organisms and has MCA equal to $\frac 1 2$.  We proceed to compute its expected value in competition with species A.  
$$E[B, A] = \left( \frac 1 2 + \eps_0\right) \left( \sum_{k \leq j} n_A (x_k) - \sum_{k > j+1} n_A (x_k) \right) + \left( \frac 1 2 - \eps_1 \right) \left( \sum_{k<M-j} n_A (x_k) - \sum_{k>M-j} n_A (x_k) \right).$$
Here, it is convenient to recall the notation \eqref{knotation}, as we will also use the symmetry $n_k = n_{-k}$ to assess the expected value.  So, in the notation of \eqref{knotation}, 
$$E[B,A] = \left( \frac 1 2 + \eps_0\right) \left( \sum_{k=\frac M 2 -j} ^{\frac M 2} n_k - \sum_{k=1} ^{\frac M 2 - j - 2} n_k - n_0 - \sum_{k=1} ^{\frac M 2} n_k \right)$$
$$+ \left( \frac 1 2 - \eps_1\right) \left( \sum_{k=1} ^{\frac M 2} n_k + n_0 + \sum_{k=1} ^{\frac M 2 - j - 1} n_k - \sum_{k= \frac M 2 - j + 1} ^{\frac M 2} n_k \right).$$
We can simplify 
$$\sum_{k=\frac M 2 -j} ^{\frac M 2} n_k - \sum_{k=1} ^{\frac M 2 - j - 2} n_k - n_0 - \sum_{k=1} ^{\frac M 2} n_k  = 
- n_0 - \sum_{k=1} ^{\frac M 2 - j - 2} n_k- \sum_{k=1} ^{\frac M 2 -j -1} n_k,$$
as well as
$$\sum_{k=1} ^{\frac M 2} n_k + n_0 + \sum_{k=1} ^{\frac M 2 - j - 1} n_k - \sum_{k= \frac M 2 - j + 1} ^{\frac M 2} n_k = n_0 + \sum_{k=1} ^{\frac M 2 -j-1} n_k + \sum_{k=1} ^{\frac M 2 -j} n_k$$
$$=  n_{\frac M 2 - j} + n_{\frac M 2 -j-1} + n_0 + \sum_{k=1} ^{\frac M 2 -j-1} n_k + \sum_{k=1} ^{\frac M 2 - j - 2} n_k.$$
Hence, 
$$E[B, A] = \left( \frac 1 2 - \eps_1 - \frac 1 2 - \eps_0 \right) \left( n_0 + \sum_{k=1} ^{\frac M 2 -j-1} n_k + \sum_{k=1} ^{\frac M 2 - j - 2} n_k \right) + \left( \frac 1 2 - \eps_1 \right) \left( n_{\frac M 2 - j} + n_{\frac M 2 - j - 1} \right).$$
We note that by the definition of \eqref{knotation} and by the symmetry of the SBD of A, 
$$n_{\frac M 2 - j} = n_A \left( \frac 1 2 + \frac 1 M \left( \frac M 2 - j \right)\right) = n_A \left( 1 - \frac j M \right) = n_A \left( \frac j M \right) = n_A (x_j).$$
Similarly, 
$$n_{\frac M 2 - j - 1} = n_A \left( \frac 1 2 + \frac{j+1}{M} \right) = n_A \left( \frac{j+1}{M}\right) = n_A (x_{j+1}).$$
By the definition of $n_k$ in \eqref{knotation} we also see that 
$$n_0 + \sum_{k=1} ^{\frac M 2 -j-1} n_k + \sum_{k=1} ^{\frac M 2 - j - 2} n_k = \sum_{k=j+1} ^{M-j-2} n_A (x_k).$$
So, we have computed 
$$E[B,A] = - (\eps_0 + \eps_1) \left(  \sum_{k=j+1} ^{M-j-2} n_A (x_k) \right) + \left( \frac 1 2 - \eps_1 \right) \left( n_A (x_j) + n_A (x_{j+1}) \right).$$
We see that setting $j=0$, we obtain the same result as in that particular case.  It is essentially the same argument for $j=0, 1, \ldots, \frac M 2 - 2$, however, it is useful for applications involving studying particular SBDs to complete the proof for all cases $j=0, 1, \ldots, \frac M 2 - 2$.  By the assumption \eqref{jmax} on $j$, and since A is not the uniform SBD, 
\begin{equation} \label{notuj} (M-2j-2) n_A (x_j) \geq  \sum_{k=j+1} ^{M-j-2} n_A (x_k) \implies (M-2j-1) n_A (x_j) \geq  n_A (x_j) + \sum_{k=j+1} ^{M-j-2} n_A (x_k). \end{equation} 
At this point, we recall the definition of $\eps_1$, \eqref{eps1j}, computing 
$$\eps_0 + \eps_1 = \eps_0 + \frac{1 + (2+2j) \eps_0}{2(M-j)} = \frac{1 + 2(M-j) \eps_0 + (2+2j)\eps_0}{2(M-j)} = \frac{1 + 2(M+1)\eps_0}{2(M-j)}.$$
We also compute 
$$\frac 1 2 - \eps_1 = \frac 1 2 -  \frac{1 + (2+2j) \eps_0}{2(M-j)} = \frac{ M-j  - 1 - (2+2j) \eps_0}{2(M-j)}.$$
Hence, we see that 
$$E[B,A] = - \frac{1 + 2(M+1)\eps_0}{2(M-j)} \left(  \sum_{k=j+1} ^{M-j-2} n_A (x_k) \right) + \frac{ M-j  - 1 - (2+2j) \eps_0}{2(M-j)} \left( n_A (x_j) + n_A (x_{j+1}) \right).$$
Since $M>j$, so $M - j > 0$, we see that 
$$E[B,A] > 0 \iff $$
$$ \left( 1 + 2(M+1)\eps_0 \right) \left(  \sum_{k=j+1} ^{M-j-2} n_A (x_k) \right) < \left(M-j  - 1 - (2+2j)\eps_0 \right) \left( n_A (x_j) + n_A (x_{j+1}) \right)$$
$$\iff  \left( 1 + 2(M+1)\eps_0 \right) \left(  \sum_{k=j+1} ^{M-j-2} n_A (x_k) \right) + (2+2j)\eps_0 \left( n_A (x_j) + n_A (x_{j+1}) \right) $$
$$< ( M-j - 1) \left( n_A (x_j) + n_A (x_{j+1}) \right).$$
It is convenient to group the terms with $\eps_0$ all together, so we must show that 
$$ \sum_{k=j+1} ^{M-j-2} n_A (x_k) + \eps_0 \left[ 2(M+1)  \left( \sum_{k=j+1} ^{M-j-2} n_A (x_k) \right) + (2+2j) \left( n_A (x_j) + n_A (x_{j+1}) \right) \right] $$
$$< ( M-j - 1) \left( n_A (x_j) + n_A (x_{j+1}) \right).$$

By \eqref{notuj}, and since $j>0$, 
$$(M- j - 1) \left( n_A (x_j) + n_A (x_{j+1}) \right) > (M-2j - 1) n_A (x_j) \geq  n_A (x_j) +  \sum_{k=j+1} ^{M-j-2} n_A (x_k).$$
We therefore see that 
$$(M-j - 1) \left( n_A (x_j) + n_A (x_{j+1}) \right) > n_A (x_j) +  \sum_{k=j+1} ^{M-j-2} n_A (x_k).$$
Hence, it is enough to prove that we can choose $\eps_0 > 0$ sufficiently small so that 
$$ \eps_0 \left[ 2(M+1) \sum_{k=j+1} ^{M-j-2} n_A (x_k) + (2+2j) \left( n_A (x_j) + n_A (x_{j+1}) \right) \right] < n_A(x_j).$$
Well, this can be achieved by simply demanding that 
$$0<\eps_0 < \frac{n_A (x_j)}{\left( 2(M+1) \sum_{k=j+1} ^{M-j-2} n_A (x_k) + (2+2j) \left( n_A (x_j) + n_A (x_{j+1}) \right) \right) }.$$

Finally, we consider the last remaining cases.  We still have the assumptions that A is not the uniform SBD, and that its SBD is symmetric about $\frac 1 2$.  In these cases we have $j$ defined as in \eqref{jmax} satisfying 
$$j = \frac M 2 - 1 \textrm{ or } j = \frac M 2.$$
We first handle the case $j=\frac M 2$.  When this is the case, by the definition of \eqref{jmax}, using \eqref{knotation}, we note that we have the inequality 
\begin{equation} \label{n02} n_A \left( \frac 1 2 \right) = n_0 > n_2 = n_{-2} = n_A \left( \frac 1 2 \pm \frac 2 M \right). \end{equation} 
Then, we define the species $B$ to have 
\begin{equation} \label{middle} n_B (x_j) = \begin{cases} 0 & j \leq \frac M 2 - 3 \\ 
0 & j \in \left\{ \frac M 2 - 1, \, \frac M 2, \, \frac M 2 + 2 \right\} \\
\frac{N}{3} & j= \frac M 2 - 2 \\ \frac{2 N}{3} & j = \frac M 2 + 1 \\ 0 & j \geq \frac M 2 + 3.  \end{cases} \end{equation} 
We calculate the MCA of species B:
$$\frac 1 3 \left( \frac M {2M} - \frac{2}{M} \right) + \frac 2 3 \left( \frac M {2M} + \frac 1 M \right) = \frac 1 2.$$
Since the distribution of $A$ is symmetric about the point $\frac 1 2$, we compute the expected value (using the notation \eqref{knotation}) 
$$E[B, A] = \frac 1 3 \left( -n_{-1} - n_0 - n_1 - n_2 \right) + \frac 2 3 \left( n_0 + n_{-1} + n_{-2} - n_2 \right) = \frac{n_0 - n_{2}}{3} > 0.$$
Above, we have used \eqref{n02}.  

In the very last case, we have $j$ defined in \eqref{jmax} with 
$$j = \frac M 2 - 1.$$
Then, by the definition of $j$, the symmetry of the distribution of A, and using the notation \eqref{knotation}, 
\begin{equation} \label{n1big} n_1 = n_{-1} = n_A \left( \frac 1 2 \pm \frac 1 M \right) > n_{\pm 2}, \quad n_1 > n_{\pm 3}. \end{equation} 
In this case, we shall exploit this by defining 
\begin{equation} \label{n-1} n_B (x_j) = \begin{cases} 0 & j \leq \frac M 2 - 4 \\ \frac N 4 & j = \frac M 2 - 3 \\ 0 & j \in \left\{ \frac M 2 - 2, \, \frac M 2 \, \frac M 2 + 1, \, \frac M 2 + 3 \right\} \\ \frac N 4 & j = \frac M 2 - 1 \\ \frac N 2 & j = \frac M 2 + 2 \\ 0 & j \geq \frac M 2 + 4. \end{cases} \end{equation} 
We compute the MCA:
$$\frac 1 4 \left( \frac{M}{2M} - \frac 3 M \right) + \frac 1 4 \left( \frac{M}{2M} - \frac 1 M \right) + \frac 2 4 \left( \frac{M}{2M} + \frac 2 M \right) = \frac 1 2.$$
We compute the expected value 
$$E[B, A] = \frac 1 4 \left( -n_{-2} - n_{-1} - n_0 -n_1 - n_2 - n_3 \right) + \frac 1 4 \left( n_{-2} + n_{-3} - n_0 -n_1 -n_2 - n_3 \right) $$
$$+ \frac 2 4 \left( n_{-3} + n_{-2} + n_{-1} + n_0 + n_1 - n_3 \right)$$
$$= \frac 1 4 \left( -2n_2 - 2n_1 - n_0 - n_3 \right) + \frac 1 4 \left( -n_0 - n_1 \right) + \frac 2 4 \left( n_2 + 2n_1 + n_0 \right)$$
$$=\frac 1 4 \left( n_1 - n_3 \right) > 0.$$
We have the above inequality from \eqref{n1big}.  

Consequently, if $A$ does not have the uniform SBD, we have shown how to construct a species $B$ with $\mca(B) \leq \frac 1 2$, and with $N_B \leq N_A$, such that 
$$E[B,A] >0.$$  
As we have computed in \eqref{0sum}, this implies that 
$$E[A,B] = - E[B,A] < 0.$$ 
Consequently, 
$$E[B,A] > 0 > E[A,B] = - E[B,A].$$
Hence no such SBD, $A$, is an NES.  Since all ESS are NES, no such SBD is an ESS.  Moreover, the uniform SBD is also not an ESS, because 
$$E[U,U] = 0 = E[A,U] = E[A,A] \textrm{ for all SBDs, $A$, with } \mca(A) = \frac 1 2.$$
However, we have proven in the proposition that for all SBDs with MCA less than or equal to $\frac 1 2$, 
$$E[U,A] \geq 0 \geq E[A,U] = - E[U,A].$$
Hence, $U$ is the unique NES.  \qed
\qed

\end{appendix}

%

 
 
\end{document}